\documentclass{emulateapj}

\newcommand{\be}{\begin{equation}}
\newcommand{\ee}{\end{equation}}
\newcommand{\hompc}{\,h\,{\rm Mpc}^{-1}}
\newcommand{\mpcoh}{\,h^{-1}\,{\rm Mpc}}
\newcommand{\gs}{\mathrel{\lower0.6ex\hbox{$\buildrel {\textstyle >}
 \over {\scriptstyle \sim}$}}}
\newcommand{\ls}{\mathrel{\lower0.6ex\hbox{$\buildrel {\textstyle <}
 \over {\scriptstyle \sim}$}}}

\shorttitle{Measuring the matter density using baryon oscillations in the SDSS}
\shortauthors{Percival et al.}

\begin{document}

\title{Measuring the matter density using baryon oscillations in the SDSS}

\author{
Will J.\ Percival\altaffilmark{1},
Robert C.\ Nichol\altaffilmark{1}, 
Daniel J.\ Eisenstein\altaffilmark{2},
David H.\ Weinberg\altaffilmark{3},
Masataka Fukugita\altaffilmark{4},
Adrian C.\ Pope\altaffilmark{5,6},
Donald P.\ Schneider\altaffilmark{7},
Alex S.\ Szalay\altaffilmark{5},
Michael S. Vogeley\altaffilmark{8},
Idit Zehavi\altaffilmark{9},
Neta A.\ Bahcall\altaffilmark{10},
Jon Brinkmann\altaffilmark{11},
Andrew J.\ Connolly\altaffilmark{12},
Jon Loveday\altaffilmark{13},
Avery Meiksin\altaffilmark{14}
}

\begin{abstract}

  We measure the cosmological matter density by observing the
  positions of baryon acoustic oscillations in the clustering of
  galaxies in the Sloan Digital Sky Survey (SDSS). We jointly analyse
  the main galaxies and LRGs in the SDSS DR5 sample, using over half a
  million galaxies in total. The oscillations are detected with
  99.74\% confidence (3.0$\sigma$ assuming Gaussianity) compared to a
  smooth power spectrum. When combined with the observed scale of the
  peaks within the CMB, we find a best-fit value of
  $\Omega_M=0.256^{+0.029}_{-0.024}$ (68\% confidence interval), for a
  flat $\Lambda$ cosmology when marginalising over the Hubble
  parameter and the baryon density. This value of the matter density
  is derived from the locations of the baryon oscillations in the
  galaxy power spectrum and in the CMB, and does not include any
  information from the overall shape of the power spectra. This is an
  extremely clean cosmological measurement as the physics of the
  baryon acoustic oscillation production is well understood, and the
  positions of the oscillations are expected to be independent of
  systematics such as galaxy bias.

\end{abstract}


\altaffiltext{1}{Institute of Cosmology and Gravitation, Mercantile
  House, Hampshire Terrace, University of Portsmouth, Portsmouth, P01
  2EG, UK}
\altaffiltext{2}{Steward Observatory, University of Arizona, 933
  N. Cherry Ave., Tucson, AZ 85121, USA}
\altaffiltext{3}{Department of Astronomy, The Ohio State University,
  Columbus, OH 43210, USA}
\altaffiltext{4}{Institute for Cosmic Ray Research, University of Tokyo, 
  Kashiwa 277-8582, Japan}
\altaffiltext{5}{Department of Physics and Astronomy,
  The Johns Hopkins University,
  3701 San Martin Drive, Baltimore, MD 21218, USA}
\altaffiltext{6}{Institute for Astronomy, University of Hawaii, 
    2680 Woodlawn road, Honolulu, HI 96822, USA}
\altaffiltext{7}{Department of Astronomy and Astrophysics, The 
  Pennsylvania State University, University Park, PA 16802, USA}
\altaffiltext{8}{Department of Physics, Drexel University, Philadelphia, 
  PA 19104, USA}
\altaffiltext{9}{Department of Astronomy, Case Western Reserve University, 
  Cleveland, OH 44106, USA}
\altaffiltext{10}{Department of Astrophysical Sciences, Princeton University,
  Princeton, NJ 08544, USA}
\altaffiltext{11}{Apache Point Observatory, P.O. Box 59, Sunspot, NM 88349, USA}
\altaffiltext{12}{Department of Physics and Astronomy, University of 
    Pittsburgh, Pittsburgh, PA 15260, USA}
\altaffiltext{13}{Astronomy Centre, University of Sussex, Falmer, Brighton, 
  BN1 9QH, UK}
\altaffiltext{14}{SUPA; Institute for Astronomy, University of Edinburgh,
  Royal Observatory, Blackford Hill, Edinburgh, EH9 3HJ, UK}

\section{INTRODUCTION}

Baryonic Acoustic Oscillations (BAO) are predicted in the matter
distribution with a calibration that depends on $\Omega_M h^2$
\citep{silk68,peebles70,sunyaev70,bond84,bond87,holtzman89}.  The
oscillations arise because sound waves in the coupled baryon-photon
plasma after an inflationary epoch will lead to the expansion of the
baryonic material in a spherical shell around a small perturbation,
reaching a radius $r_S(z_*)$, the comoving sound horizon size at
recombination, before sound waves are no longer supported within the
plasma \citep{bashinsky01,bashinsky02}. At the high redshifts of
interest the vacuum energy can be neglected, and $r_S(z_*)$ can be
simply written \citep{hu95}
\be
  r_S(z_*) / \mpcoh \equiv \frac{1}{100 \Omega_m^{1/2}} 
    \int_0^{a_*} \frac{c_S}{(a + a_{\rm eq})^{1/2}} \, da.
  \label{eq:rs} 
\ee
The expansion factor $a \equiv (1+z)^{-1}$ and $a_*, a_{\rm eq}$ are
the values at recombination and matter-radiation equality
respectively. Thus $r_S(z_*)$ depends on the matter density $\Omega_M$
through the expansion rate and the recombination redshift. For a
baryon density $\Omega_bh^2 \simeq 0.02$, we can approximate $c_S
\simeq 0.90 \, c / \sqrt{3}$.  Inserting $z_* = 1100$ and $a_{\rm eq}
= (23900 \,\Omega_Mh^2)^{-1}$ gives $r_S(z_*)=109\mpcoh$ for $\Omega_M=0.24$
and $h=0.73$.

In real space this leads to a peak in the correlation function at
$r_S(z_*)$. In Fourier space, this process leads to oscillations in
the power spectrum in the same way that the transform of a top-hat
function yields a sinc function. The wavelength of these oscillations
for $\Omega_M=0.24$ and $h=0.73$ is
$k_S=2\pi/109=0.06\hompc$. Numerical simulations have shown that a
number of subtle corrections are required to this simple picture,
although these corrections do not significantly affect the underlying
important physics. On large scales, there is a phase shift in the
position of the oscillations due to the contribution of the baryons to
the drag epoch \citep{eisenstein98}. We must also consider the damping
of the oscillations on small scales at high redshift \citep{silk68},
and, when modelling the oscillations at low redshifts, due to
structure formation \citep{eisenstein06}.

The resulting acoustic peaks in the cosmic microwave background (CMB)
have now been observed with extreme precision
\citep{hinshaw06,spergel06}. However, interpreting the observed
angular separation of these peaks in terms of the physics of the early
universe requires knowledge of the angular diameter distance to the
last scattering surface -- in the context of flat $\Lambda$
cosmological models this leads to a parameter degeneracy between the
matter density $\Omega_M$ and the Hubble constant $h$, and models with
the same value of $\Omega_M^{0.275}h$ have the same projected acoustic
horizon scale \citep{percival02,page03,spergel06} (throughout this
paper, a ``flat $\Lambda$'' cosmological model implies a universe that
is spatially flat with a time-independent ``dark energy'' component,
i.e., a cosmological constant). Theoretically, we should expect the
oscillations to survive in the galaxy power spectrum
\citep{meiksin99,springel05,seo05,white05,eisenstein06}, and the
effects of baryons have been previously detected on large scales in
the clustering of galaxies
\citep{percival01,miller01,cole05,eisenstein05,huetsi06}. These low
redshift observations have a different dependence on the cosmological
distance-redshift relation compared with the CMB because of the
different angular and radial projection of these features. By
comparing the two observations we can therefore probe the cosmological
expansion history in addition to the physics of the BAO production. In
this paper we use the BAO measured in the SDSS DR5 galaxy sample to
set tight, clean constraints on the cosmological matter density.

\section{MEASURING THE SDSS POWER SPECTRUM}  
\label{sec:method}

The data and method used for calculating the redshift-space power
spectrum of the latest SDSS sample are described in detail in
\citet{percival06}. In this section we summarise this information
paying particular attention to important details for the BAO
detection. The SDSS
\citep{york00,adelman06,blanton03,fukugita96,gunn98,gunn06,hogg01,ivezic04,pier03,smith02,stoughton02,tucker06}
Data Release 5 (DR5) sample represents the largest volume of the
Universe that has been mapped to date. In total we have 522280 galaxy
spectra, with 465789 of those spectra being main galaxies
\citep{strauss02} selected to a limiting magnitude $r<17.77$, or
$r<17.5$ in a small subset of the early data from the survey. The
remaining 56491 galaxies are Luminous Red Galaxies (LRGs;
\citealt{eisenstein01}), which form an extension to the survey to
higher redshifts $0.3<z<0.5$; this extension covers most of the volume
mapped. Of the main galaxies, 21310 are also classified as LRGs, so
our sample includes 77801 LRGs in total. In this paper, and its
companion \citep{percival06}, we analyse the combined sample of main
galaxies and LRGs thereby including correlations between the two
samples. Although the main galaxy sample contains significantly more
galaxies than the LRG sample, the LRG sample covers more volume, and
therefore contains almost all of the cosmological
information. However, as shown in figure~8 of \citet{percival06}, the
contribution of pairs of galaxies where one galaxy is a main sample
galaxies and one is a LRG are not negligible compared with LRG--LRG
pairs, justifying the added complexity of an analysis using both data
sets.

The sample now contains 60\% more LRGs than considered in the first
measurement of BAO in the SDSS LRG sample \citep{eisenstein05}; with
the increased precision due to this increase in volume and the
main--LRG pairs of galaxies, we can now attempt to derive constraints
on the matter density that do not rely on the overall shape of the
Cold Dark Matter (CDM) power spectrum but only on the peak locations
relative to a smooth underlying spectrum. Although ignoring the
overall shape removes information, we gain in robustness due to the
increased simplicity of the physics producing the cosmological
constraint; measurements of the matter density from the overall shape
of the galaxy power spectrum are potentially affected by galaxy bias
\citep{percival06}, or the form of the fluctuation spectrum from the
inflationary model.

The redshift-space clustering power spectrum of this sample has been
calculated using a Fourier based technique \citep{FKP,PVP}. This
method uses a simple model for the relative bias of galaxies to remove
effects in the power spectrum due to pairs of galaxies with different
expected clustering amplitudes. The power spectrum is recalculated for
31 flat $\Lambda$ cosmological models, with matter density
$0.1\le\Omega_M\le0.4$ and $\Delta\Omega_M=0.01$. For each of these
cosmological models, we have created 2000 log-normal catalogues (using
the method described in \citealt{cole05}) with power spectra
calculated using a linear CDM model with parameters chosen to
approximately match the amplitude and shape of the recovered power for
$0.01<k<0.15\hompc$. These power spectra are used to calculate a
covariance matrix for each model, although we then minimise the effect
of Monte-Carlo noise in each element in the set of covariance matrices
as a function of $\Omega_M$ by smoothing using a 4-node cubic spline
with nodes at $\Omega_M=0.1,0.2,0.3,0.4$. This ensures a smooth
progression in the error estimation over the set of cosmological
models. The convolving effect of the survey geometry on the power
spectrum has also been quantified by spline fitting the Fourier
transform of the window function. This fit is used to smooth all
models before comparison with the data.

\section{MODELLING THE BARYON ACOUSTIC OSCILLATIONS}

We model the true galaxy power spectrum on scales $0<k<0.3\hompc$ with
a two component model. The overall shape was matched using a cubic
spline fit with 8 nodes separated by $\Delta k=0.05\hompc$ and
$0.025\le k\le0.375\hompc$, and an additional node at
$k=0.001\hompc$. This smooth model was then modulated by a higher
frequency component, constructed as follows: the sinusoidal BAO term
in a standard CDM transfer function was estimated for the parameters
to be tested from numerical fits \citep{eisenstein98,eisenstein06},
including a damping term to approximately correct for non-linear
structure formation, and the multiplicative effect on a CDM power
spectrum was isolated. This multiplicative term was then applied to
the smooth cubic spline power spectrum rather than a CDM model. The
model power spectra are adjusted for the effects of the survey
geometry by convolving with the appropriate window function and
correcting for our lack of knowledge about the true mean density of
galaxies by subtracting a multiple of the Fourier window function from
the model power so that $P(0)=0$ (see \citealt{percival06} for
details).

This procedure separates the physics of the BAO from that governing
the overall shape of the power spectrum, including both cosmological
and galaxy formation effects, and ensures that the cosmological
constraints presented in this paper only come from the BAO and not
from the additional physics encoded in the power spectrum. In
particular, forming a model power spectrum in this way allows for
non-linear effects and galaxy bias to change the overall shape of the
power spectrum, and damp the oscillations on small scales. The model
does not allow the BAO scale to change, although it is worth
emphasising that the model can lead to an apparent change in the
positions of the peaks and troughs in the model power spectra, caused
by the addition of a tilted smooth power spectrum component (for
example the 1-halo term in the halo model), or multiplication by such
a smooth component. Recent theoretical models of BAO in galaxy power
spectra find such an apparent change in the positions of the peaks and
troughs \citep{seo05,smith06}, but have not shown evidence for a
change in the BAO scale. Because we split into the BAO and a smooth
component, any such apparent shift in the observed BAO scale is
removed by our analysis method. Extra multiplicative or additive low
frequency power spectrum components will change the BAO damping, which
will be a function of the luminosity of the galaxy sample. For a
combined sample of galaxies with different luminosity such as analysed
in this paper, the form and amplitude of the small-scale BAO damping
will depend on the details of the galaxy sample. However, the damping
does not have a significant effect on our results, so our results are
expected to be robust to such complexities.

Because of the large volume observed, the Fourier modes are not
strongly correlated -- correlations between modes drop to $<0.33$ for
$\Delta k>0.01\hompc$. The correlation scale is therefore
significantly smaller than the BAO wavelength. However, we calculate
the likelihood of the data given each model to be tested assuming that
the data are drawn from a multi-variate Gaussian distribution with
covariance matrix calculated as described in Section~\ref{sec:method}
(including the effect of the change in the determinant as a function
of $\Omega_M$). For each cosmological model tested, we calculate the
maximum likelihood obtained after varying the values of the smooth
spline function at the nine nodes for models where we include or
exclude the BAO model. Ideally, excluding the BAO, we would find that
the likelihood does not change with the cosmological model. However,
because the data change with $\Omega_M$ we find small variations in
the likelihood even when only fitting a smooth curve to the power
spectra. To remove this ``noise'', we only consider the ratio of the
different likelihoods including and excluding the BAO signal.

\begin{figure}[tb]
  \plotone{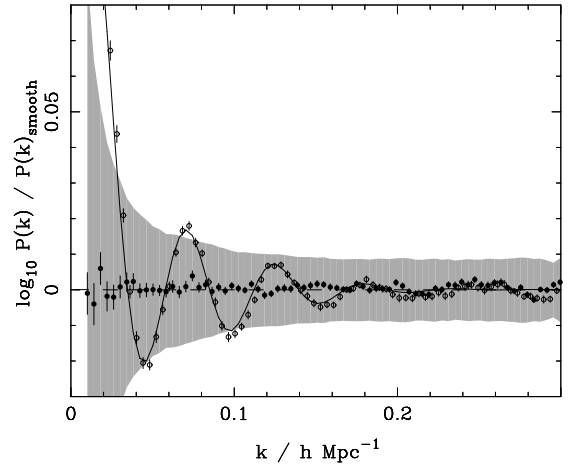}
  \caption{Results from fitting to 200 mock linear CDM power spectra
    calculated assuming $h=0.73$, $\Omega_M=0.24$ and $n_s=0.96$, with
    our two component spline + BAO model. We plot the average ratio
    between input power and spline fit from 100 mock CDM power spectra
    with no baryon oscillations (open circles with 1$\sigma$ errors),
    and 100 mocks with baryon oscillations assuming
    $\Omega_b/\Omega_M=0.174$ (solid circles with 1$\sigma$
    errors). The expected residual in each case is shown by the solid
    lines.  From these fits, we find that the difference between input
    and recovered power $\Delta P(k)/P(k)<0.015$ for
    $0.01<k<0.3\hompc$, a level well below the current experimental
    error (grey shaded region). \label{fig:test_bao_plot_k0_25}}
\end{figure}

This method relies on the spline curve being able to fit the power
spectrum shape. In order to test this, we have fitted two sets of 100
linear CDM power spectra (calculated using the fits of
\citealt{eisenstein98} assuming $h=0.73$, $\Omega_M=0.24$ and
$n_s=0.96$) with our combined spline+BAO model. The first set of mock
data contained BAO with $\Omega_b/\Omega_M=0.174$, while the second
set had no BAO ($\Omega_b=0$). Noise was added to the mock power
spectra drawn from a multi-variate Gaussian distribution with
covariance matrix matched to the real data, and the power spectra were
convolved with the window function of the survey (for a cosmological
distance model with $\Omega_M=0.24$). For the combined spline+BAO
model that we fit to these data, we have assumed that there is no
small scale damping, and have fixed the cosmological parameters at the
input values. The spline curves are then allowed to vary to fit the
data, and the average residuals between the mock power and the spline
fit are plotted in Fig.~\ref{fig:test_bao_plot_k0_25}, compared to the
expected residuals. The results match those expected at a level well
below the error on the power spectrum; this, and the acceptable
average $\chi^2$ values of the fits (64.2 and 66.2 for the sets of
models with and without BAO, given 64 degrees-of-freedom) show that
the spline+BAO model can match the features expected in a linear CDM
power spectrum.

Finally, we end this section with a brief discussion on the node
separation chosen for the spline fit. If too many nodes are chosen,
then the spline fit can itself match the BAO, leading to a small
likelihood ratio between models with and without BAO even if BAO are
strong in the data. Conversely, if too few nodes are chosen, the
addition of BAO to the model will not necessarily match high-frequency
features in the observed power spectrum, and might instead simply help
to fit the overall shape. The node spacing adopted in this work was
carefully chosen based on the analysis of mock power spectra, and from
the results of the fits to the SDSS power spectra -- with the chosen
separation no evidence was observed for low frequency residuals from
any of the SDSS power spectra fits, with or without BAO.

\section{RESULTS}

\begin{figure}[tb]
  \plotone{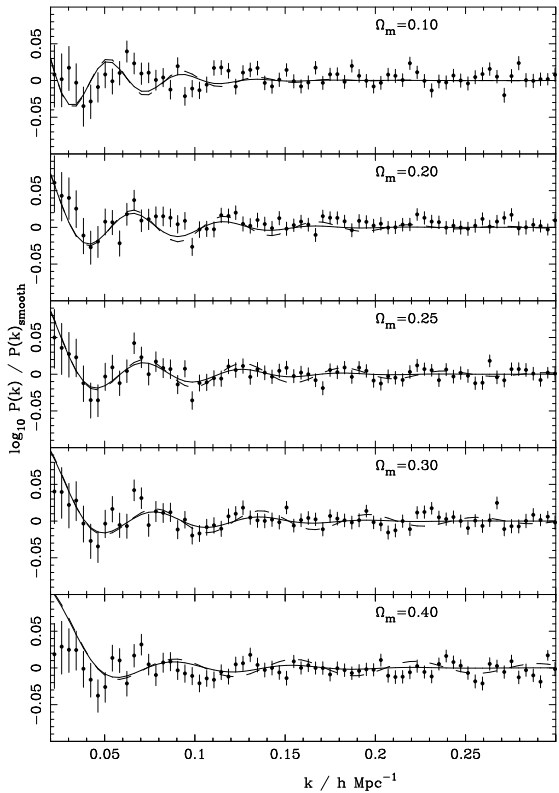}
  \caption{The ratio of the power spectra calculated from the SDSS to
    the smooth cubic spline fit that we use to model the overall shape
    of the measured power spectra (solid circles with 1-$\sigma$
    errors). Data are plotted using five flat $\Lambda$ cosmological
    models to convert from redshift to comoving distance, with matter
    densities given in each panel. For comparison, in each panel we
    also plot the BAO predicted by a CDM model with the same matter
    density, $h=0.73$, and a 17\% baryon fraction (solid lines). The
    dashed lines show the same models without the low-redshift
    small-scale damping term. As can be seen, the observed
    oscillations approximately match those predicted by this model for
    $0.2\le\Omega_M\le0.3$, but fail for higher or lower matter
    densities. \label{fig:pk_bao_omm}}
\end{figure}

In Fig.~\ref{fig:pk_bao_omm}, we plot measured power spectra
determined assuming 5 different values of the matter density, divided
by the best-fit smooth cubic spline fit (solid circles). The spline
fits were calculated after fitting the data including a possible BAO
contribution with fixed $\Omega_b/\Omega_M\sim0.17$ and $h=0.73$, and
the appropriate $\Omega_M$. These data are compared with the BAO
model, and show that the model and data match only if
$0.2\ls\Omega_M\ls0.3$, assuming $\Omega_b/\Omega_M\sim0.17$ and
$h=0.73$. For $\Omega_M=0.26$, the baryon oscillations are required
with a likelihood ratio $2\ln{\cal L}=9.02$, corresponding to a
99.74\% confidence of detection (3.0$\sigma$ assuming Gaussianity).

\begin{figure}[tb]
  \plotone{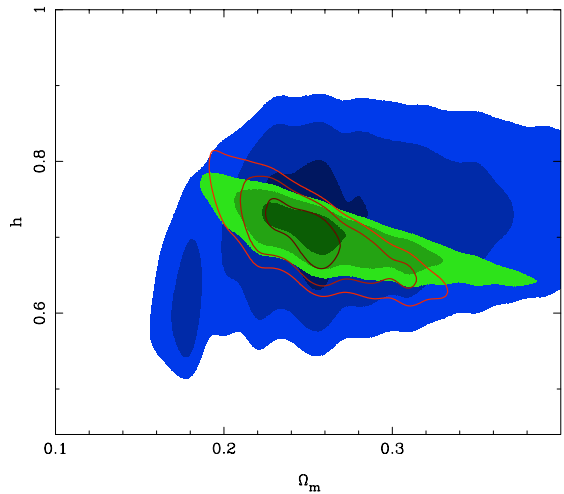}
  \caption{Likelihood contours in the $h-\Omega_M$ plane, derived from
    measurements of the BAO observed in the SDSS combined with
    constraints from other cosmological data. The intensities are
    separated by $-2\ln{\cal L}=2.3,\,6.0,\,9.2$, corresponding to
    two-parameter confidence of 68\%, 95\% and 99\% for a Gaussian
    distribution.  The blue shaded region shows the constraints for
    flat $\Lambda$ cosmologies combining the SDSS BAO data with a low
    redshift constraint on $h=0.72\pm0.08$ for flat
    $\Lambda$-cosmologies. The green region shows the combination of
    the SDSS constraint with the 3-year WMAP constraints on
    $\Omega_M^{0.275}h$ and $\Omega_bh^2$, again for flat
    $\Lambda$-cosmologies. The overlaid red contours were calculated
    by instead combining with a constraint on $\Omega_Mh^2$ from the
    peak heights in the CMB together with the constraint on
    $\Omega_bh^2$. This relaxes the assumption of a flat
    $\Lambda$-cosmology, by removing the dependence on the distance to
    the last scattering surface. There is still a dependence on the
    comoving distance-redshift relation over the survey, but
    observations of type Ia supernovae constrain this to be close to
    that expected for a flat $\Lambda$-cosmology. The degeneracies in
    the $h-\Omega_M$ plane induced by different cosmological
    observations are discussed at length in
    \citet{tegmark06}. \label{fig:like_bao_om_vs_h}}
\end{figure}

Because the amplitude of the BAO depends on the fraction of matter
that is baryonic, the SDSS observations constrain the baryon fraction,
although this constraint is weak compared to constraints on
$\Omega_bh^2$ from CMB observations. The scale of the BAO depends on
$\Omega_M$ and $h$, so we need one other piece of information about a
combination of these parameters to break this degeneracy and measure
$\Omega_M$. We consider three options, yielding the three sets of
likelihood contours in the $h-\Omega_M$ plane that are shown in
Fig.~\ref{fig:like_bao_om_vs_h}. First, we combine our low redshift
BAO measurement with the HST Key Project \citep{freedman01} estimate
of the Hubble parameter $h=0.72 \pm 0.08$ ($1\sigma$ errors).  This
gives $\Omega_M = 0.256^{+0.049}_{-0.029}$ (all the error bars quoted
in this paper span the 68\% confidence interval).  Here we have
marginalized over the uncertainty in $h$ (assumed to be Gaussian) and
marginalised over a uniform prior on the baryon density
$0.008<\Omega_bh^2<0.034$. This estimate of $\Omega_M$ assumes a flat
$\Lambda$ cosmological model, but dark energy and spatial curvature
affect the result only through their influence on the
distance-redshift relation for the galaxies within our
sample. Measurements of Type Ia supernovae already demonstrate that
this relation is close to that expected for a flat $\Lambda$ universe;
we estimate that the allowable residual effect on $\Omega_M$ is at
most $\pm 0.02$.

As an alternative to direct $h$ measurement, we consider the
combination of our low-$z$, redshift-space measurement of the BAO
scale with the angular scale of the acoustic oscillations measured in
the CMB. The 3-year WMAP data \citep{spergel06} yields the constraint
$\Omega_M^{0.275}h = 0.492^{+0.008}_{-0.017}$ after marginalizing over
the scalar spectral index and the baryon density. Combined with our
BAO measurement, this yields $\Omega_M = 0.256^{+0.029}_{-0.024}$, and
$h = 0.709^{+0.022}_{-0.027}$. For simplicity, we have made the good
but not perfect approximation that the WMAP baryon density constraint,
$100\Omega_b h^2 = 2.233^{+0.072}_{-0.091}$, is independent of the
acoustic scale constraint. Relaxing the baryon density constraint by a
factor of four makes essentially no difference to either the best-fit
$\Omega_M$ or the error bars.  This second determination of $\Omega_M$
relies more heavily on the assumption of a flat $\Lambda$ universe,
since space curvature or a different dark energy component would
change the distance to the last scattering surface and hence the
angular scale of the CMB acoustic peaks.

We can weaken the dependence on the assumption of a flat $\Lambda$
cosmology by instead combining our low redshift BAO measurement with
the WMAP constraint on the matter density, $\Omega_M h^2 =
0.1268^{+0.0072}_{-0.0095}$. This constraint comes from the relative
{\it heights} of the CMB acoustic peaks, so the physics that underlies
it is somewhat different, and somewhat more degenerate with the
parameters of the inflationary fluctuation spectrum. The dependence on
the assumption of a flat $\Lambda$ cosmology is weaker because the
relative CMB peak heights only depend on the distance to the last
scattering surface through the weak, and predominantly large-angle,
ISW effect. Additionally, we have already argued that observations of
type Ia supernovae constrain the distance-redshift relation for the
low-redshift galaxies to be close to a flat $\Lambda$ cosmology
(leaving a residual effect $\Delta\Omega_M$ at most $\pm 0.02$). The
direction of the $\Omega_M h^2$ constraint in the $h-\Omega_M$ plane
is more complementary to the low redshift BAO observations for
measuring $\Omega_M$ compared with the CMB peak positions, and this
combination yields $\Omega_M = 0.256^{+0.020}_{-0.022}$, again
marginalising over the WMAP baryon density constraint.

\section{DISCUSSION}

In a separate paper, \citet{tegmark06} carry out full multi-parameter
fits to the LRG power spectrum and WMAP data over a broader CDM model
space, drawing on both the BAO measurement and the broadband shape of
the power spectrum.  The more focused analysis presented in this paper
is complementary, obtaining constraints with minimal dependence on
detailed cosmological assumptions.  Within the context of flat
$\Lambda$ models, the measurement of the matter density, $\Omega_M =
0.256^{+0.029}_{-0.024}$, from the locations of the acoustic
oscillations (the second combination above) is especially ``clean,''
relying on a single piece of well understood physics. In particular,
we have decoupled constraints from the BAO with constraints from the
overall shape of the power spectrum on the same scales. As
demonstrated in \citet{percival06}, the overall shape of the power
spectrum, even on scales where the matter clustering has not deviated
strongly from a linear CDM model is dependent on the galaxy population
studied. It is therefore imperative to accurately model galaxy bias
before robust cosmological constraints can be derived from such
observations. By simply considering the BAO in this paper, we avoid
this complexity.

Given the different physics probed, the consistency of the three
$\Omega_M$ estimates calculated with different additional information,
itself provides support for the assumptions of a flat universe with a
cosmological constant. With larger samples and a wider redshift range,
the BAO ``standard ruler'' can be used to test these assumptions at
high precision through the comoving distance-redshift relation
\citep{blake03,seo03}.  Forthcoming surveys have been designed exploit
this effect.  It is clear that we are entering an era where BAO offer
an extremely attractive route to strong, robust cosmological
constraints.

\acknowledgments

WJP is grateful for support from a PPARC fellowship, and RCN for a EU
Marie Curie Excellence Chair.

Funding for the SDSS and SDSS-II has been provided by the Alfred
P. Sloan Foundation, the Participating Institutions, the National
Science Foundation, the U.S. Department of Energy, the National
Aeronautics and Space Administration, the Japanese Monbukagakusho, the
Max Planck Society, and the Higher Education Funding Council for
England. The SDSS Web Site is \url{http://www.sdss.org/}.

The SDSS is managed by the Astrophysical Research Consortium for the
Participating Institutions. The Participating Institutions are the
American Museum of Natural History, Astrophysical Institute Potsdam,
University of Basel, Cambridge University, Case Western Reserve
University, University of Chicago, Drexel University, Fermilab, the
Institute for Advanced Study, the Japan Participation Group, Johns
Hopkins University, the Joint Institute for Nuclear Astrophysics, the
Kavli Institute for Particle Astrophysics and Cosmology, the Korean
Scientist Group, the Chinese Academy of Sciences (LAMOST), Los Alamos
National Laboratory, the Max-Planck-Institute for Astronomy (MPIA),
the Max-Planck-Institute for Astrophysics (MPA), New Mexico State
University, Ohio State University, University of Pittsburgh,
University of Portsmouth, Princeton University, the United States
Naval Observatory, and the University of Washington.

\setlength{\bibhang}{2.0em}

\label{lastpage}

\end{document}